\documentstyle[12pt]{article}
\newcommand{\eq}{\begin{equation}}
\newcommand{\en}{\end{equation}}
\newcommand{\eqn}{\begin{eqnarray}}
\newcommand{\enn}{\end{eqnarray}}
\newcommand{\CR}{\nonumber \\}

\newcommand{\I}{{\rm i}}
\newcommand{\pa}{\partial}

\newcommand{\A}{\alpha}
\newcommand{\B}{\beta}
\newcommand{\D}{\delta}
\newcommand{\DE}{\Delta}

\newcommand{\lm}{\lambda}

\newcommand{\bg}{{\bf g}}

\newcommand{\hbg}{\hat{\bf g}}

\newcommand{\dpsi}{\psi^{\dagger}}

\newcommand{\tB}{\tilde{\beta}}
\newcommand{\tG}{\tilde{\gamma}}
\newcommand{\tb}{\tilde{b}}
\newcommand{\tc}{\tilde{c}}
\newcommand{\tJ}{\widetilde{J}}
\newcommand{\tj}{\widetilde{j}}

\newcommand{\tH}{\widetilde{H}}

\newcommand{\tW}{\widetilde{W}}
\newcommand{\cQ}{{\cal Q}}
\newcommand{\cR}{{\cal R}}
\newcommand{\cU}{{\cal U}}
\newcommand{\cT}{{\cal T}}
\newcommand{\cG}{{\cal G}}
\newcommand{\cW}{{\cal W}}

\begin{document}
\renewcommand{\thefootnote}{\fnsymbol{footnote}}
\begin{titlepage}
\null
\begin{flushright}
UTHEP-277 \\
May 1994
\end{flushright}
\vspace{2.5cm}
\begin{center}
{\Large \bf
Lie Superalgebra and \\
Extended Topological Conformal Symmetry \\
in Non-critical $W_{3}$ Strings
\par}
\lineskip .75em
\vskip 3em
\normalsize
{\large Katsushi Ito}
\vskip 1.5em
{\it Institute of Physics, University of Tsukuba, Ibaraki 305, Japan}
\vskip 1.5em
and
\vskip 1.5em
{\large Hiroaki Kanno}
\vskip 1.5em
{\it Department of Mathematics, Hiroshima University, \\
Higashi-Hiroshima 724, Japan }
\vskip 1.5em
{\bf Abstract}
\end{center} \par

We obtain a new free field realization of $N=2$ super $W_{3}$ algebra
using the technique of quantum hamiltonian reduction.
The construction is based on a particular choice of the simple
root system of the affine Lie superalgebra $sl(3|2)^{(1)}$
associated with a non-standard $sl(2)$ embedding.
After twisting  and a similarity transformation, this $W$ algebra can
be identified as the extended topological conformal algebra of non-critical
$W_{3}$ string theory.
\end{titlepage}
\renewcommand{\thefootnote}{\arabic{footnote}}
\setcounter{footnote}{0}
\baselineskip=0.7cm

Many attempts have been done to construct non-critical string theories with
$W$ algebra symmetry (non-critical $W$ strings)\cite{Wgra}, which may be
defined beyond the $c=1$ barrier of non-critical string theory.
Physical states of non-critical $W$-strings can be characterized by
the BRST cohomology.
It is known that the BRST current has quite non-trivial structure
due to non-linearity of $W$-algebra \cite{BRS}.
It is recently understood that the BRST algebra in non-critical string
theory may be enlarged to the twisted $N=2$ superconformal algebra
{\it i.e.} topological conformal algebra\cite{GaSe}.
Bershadsky, Lerche, Nemeschansky and Warner\cite{BLNW}
found the topological $W$-symmetry in the non-critical $W_{3}$-string.
This fact would be universal in a class of non-critical $W$-string theories
and essential for investigation of their properties as
topological strings, which is a clue to a non-perturbative formulation of
string theories in higher dimensions.

Practical computations involving (topological) $W$ algebra
often become complicated due to its non-linearity.
In this paper we regard the topological $W_{N}$ symmetry
as a result of the quantum Hamiltonian reduction of
an affine Lie superalgebra $sl(N|N-\,1)^{(1)}$.
We believe that the viewpoint of
Lie superalgebra is helpful in more systematic
understanding of the algebraic structure of topological
$W$ symmetry, which is obscured by the non-linearity.
For example, the existence of the BRST current
which has completely vanishing nilpotent OPE relation
is most clearly understood from the hidden symmetry
of $sl(N|N-\,1)^{(1)}$ \cite{BLNW}.
The Lie superalgebra may also explain the origin of
the screening operators which play an essential role
in investigating the physical spectrum, especially
the problem of $W$ gravitational dressing.
This implies the Lie superalgebra $sl(N|N-\,1)^{(1)}$
is important for geometrical aspects
of the theory, such as $W$ moduli.

In a previous paper\cite{ItKa}, we have shown the relation between
the topological conformal algebra and the Lie superalgebra
$sl(2|1)^{(1)}$ using the hamiltonian reduction.
In this article we will examine the quantum hamiltonian
reduction of an affine Lie superalgebra $sl(3|2)^{(1)}$
and study a free field realization of the $N=2$ super-$W_{3}$ algebra
\cite{N2W3} relevant to the non-critical $W_{3}$-string theory.
This reduction at classical level has been discussed
in \cite{BLNW}. However, the fermionic ghosts
that they employed are not free fields due to ghost number violation
term in the $U(1)$ current, which cannot be expected from the
standard hamiltonian reduction. We will show that
by a similarity transformation the fermionic ghosts of Bershadsky
et.al. are related to the genuine free fields which naturally
appear in the quantum hamiltonian reduction.

Let us define an  affine Lie superalgebra $\hbg=sl(N|N-1)^{(1)}$.
The algebra $\hbg$ at level $K$ is generated by bosonic currents
$J_{i,j}(z)$ ($i,j=1,\ldots, N$), $J_{N+a,N+b}(z)$ ($a,b=1,\ldots, N-1$)
and fermionic currents $j_{i,N+a}(z)$, $j_{N+a,i}(z)$ ($i=1,\ldots, N,
a=1,\ldots, N-1$).
The diagonal part of the bosonic currents satisfies the super-traceless
condition:
\eq
\sum_{i=1}^{N}J_{i,i}(z)-\sum_{a=1}^{N-1}J_{N+a,N+a}(z)=0.
\en
The operator product expansions for these currents are given by
\eqn
J_{i,j}(z)J_{k,l}(w)&=&
           {K(\D_{j,k}\D_{i,l}-\D_{i,j}\D_{k,l})\over (z-w)^{2}}+
               {\D_{j,k}J_{i,l}(w)-\D_{i,l}J_{k,i}(w) \over z-w}+\cdots, \CR
J_{N+a,N+b}(z)J_{N+c,N+d}(w)&=&
           {-K(\D_{b,c}\D_{a,d}+\D_{a,b}\D_{c,d})\over (z-w)^{2}}+
{\D_{d,a}J_{N+c,N+b}(w)-\D_{b,c}J_{N+a,N+d}(w) \over z-w}
    +\cdots, \CR
J_{i,j}(z)J_{N+a,N+b}(w)&=& {-K\D_{i,j}\D_{a,b}\over (z-w)^{2}}+\cdots,\CR
J_{i,j}(z)j_{k,N+a}(w)&=&{\D_{j,k}j_{k,N+a}(w)\over z-w}+\cdots,\CR
J_{i,j}(z)j_{N+a,k}(w)&=&{-\D_{i,k}j_{N+a,j}(w)\over z-w}+\cdots,\CR
J_{N+a,N+b}(z)j_{k,N+c}(w)&=&{\D_{a,c}j_{k,N+b}(w)\over z-w}+\cdots,\CR
J_{N+a,N+b}(z)j_{N+c,k}(w)&=&{-\D_{b,c}j_{N+a,k}(w)\over z-w}+\cdots,
\CR
j_{i,N+a}(z)j_{N+b,j}(w)&=&{K\D_{i,j}\D_{a,b}\over (z-w)^{2}}+
{\D_{a,b}J_{i,j}(w)-\D_{i,j}J_{N+b,N+a}(w)\over z-w}+\cdots .
\enn
The Lie superalgebra $\bg=sl(N|N-1)$ admits several choices
of the simple root system.
In the present paper, we use the following simple root system:
$\A_{i}=e_{i}-e_{i+1}$ ($i=1,\ldots, N-1$), $\A_{N}=e_{N}-\D_{1}$,
$\A_{a+N}=\D_{a}-\D_{a+1}$ ($a=1,\ldots, N-2$),
where we introduced an orthonormal basis $e_{i}$ ($i=1,\ldots, N$) and
$\D_{a}$ ($a=1,\ldots, N-1$), which satisfies $e_{i}\cdot e_{j}=\D_{i,j}$,
$\D_{a}\cdot\D_{b}=-\D_{a,b}$.
This simple root system, which is characterized as having only
one odd root, corresponds to a non-standard $sl(2)$ embedding.
The standard $sl(2)$ embedding gives the purely
fermionic simple root system,
which corresponds to the manifestly supersymmetric $N=2$ super-$W_{N}$
algebra \cite{It}.
The bosonic currents $J_{i,j}$ ($J_{N+a,N+b}$) for $i\neq j$ ($a\neq b$)
correspond to  even roots $e_{i}-e_{j}$ ($\D_{a}-\D_{b}$).
The fermionic currents $j_{i,N+a}$ ($j_{N+a,i}$) correspond to odd roots
$e_{i}-\D_{a}$ ($\D_{a}-e_{i}$).
The positive roots are $e_{i}-e_{j}$ ($i<j$),
$\D_{a}-\D_{b}$ ($a<b$) and $e_{i}-\D_{a}$.
The Cartan currents for simple roots $H_{i}=\A_{i}\cdot H$
($i=1,\ldots, 2n $) are defined
by $H_{i}=J_{i,i}-J_{i+1,i+1}$.

Based on this algebra, we construct the reduced phase space by imposing
the constraints.
The Lie superalgebra demands the second class constraints.
In such a case it is convenient to introduce auxiliary free fermions which
change the second class constraints to the first class\cite{BeOo2}.
Let us introduce $N-1$ pairs of fermionic fields
$(\dpsi_{N+a}, \psi_{N+a})$ ($a=1,\ldots, N-1$)
with weights $(1/2+N-a,1/2-N+a)$.
The operator product expansions are given by
\eq
\psi_{N+a}(z)\dpsi_{N+b}(w)={\D_{a,b}\over z-w}+\cdots .
\en
The constraints which are consistent with the affine Lie superalgebra are
\eqn
J_{i,j}(z)&=& \D_{i,j-1}, \quad   \mbox{ for $i<j$,} \CR
j_{i,N+a}(z)&=& \D_{i,N}\dpsi_{N+a}(z), \CR
J_{N+a,N+b}(z)&=& \D_{a,b-1}+\psi_{N+a}\dpsi_{N+b}(z), \quad
\mbox{for $a<b$} .
\label{eq:const}
\enn

Let us consider the BRST quantization of this constraint system.
Introduce fermionic ghosts
$(\tc_{i,j}(z),\tb_{i,j}(z))$ ($i< j$)
and $(\tc_{N+a,N+b}(z),\tb_{N+a,N+b}(z))$ ($a<b$) for the constrains
for bosonic currents.
Bosonic ghosts $(\tG_{i,N+a}(z),\tB_{i,N+a}(z))$ for fermionic
constraints.
The operator product expansions are given by
\eqn
\tc_{i,j}(z)\tb_{k,l}(w)&=& {\D_{i,k}\D_{k,l}\over z-w}+\cdots, \CR
\tc_{N+a,N+b}(z)\tb_{N+c,N+d}(w)&=&{\D_{a,c}\D_{b,d}\over z-w}
+\cdots,  \CR
\tB_{i,N+a}(z)\tG_{j,N+b}(w)&=& {\D_{i,j}\D_{a,b}\over z-w}+\cdots .
\enn
The BRST current is defined as
\eqn
J_{BRST}(z)&=& \sum_{i<j}\tc_{i,j}(J_{i,j}-\D_{i+1,j})
+\sum_{a<b}\tc_{N+a,N+b}(J_{N+a,N+b}-\D_{a+1,b}
                              -\psi_{N+a}\dpsi_{N+b}) \CR
& &  +\sum_{i,a}\tG_{i,N+a}(j_{i,N+a}-\D_{i,N}\dpsi_{N+a}) \CR
& & -\sum_{i<k<j}\tc_{i,k}\tc_{k,j}\tb_{i,j}
   +\sum_{a<c<b}\tc_{N+a,N+c}\tc_{N+c,N+b}\tb_{N+a,N+b} \CR
& &-\sum_{i<j,a}\tc_{i,j}\tG_{j,N+a}\tB_{i,N+a}
   -\sum_{i,a<b}\tG_{i,N+a}\tc_{N+a,N+b}\tB_{i,N+b} .
\enn
Indeed the BRST charge $Q_{BRST}=\oint {d z \over 2\pi \I} J_{BRST}(z)$
satisfies the nilpotency condition: $Q_{BRST}^2=0$.
We wish to investigate the BRST cohomology\footnote{This BRST
cohomology should not be confused with the BRST cohomology
which defines the physical spectrum of non-critical strings.}
on the space of operator algebra ${\cal A}_{tot}$:
\eq
{\cal A}_{tot}=U\hbg \otimes Cl_{\psi,\dpsi}\otimes
Cl_{\tb,\tc}\otimes H_{\tB,\tG},
\en
where $U\hbg$ is the universal enveloping algebra of $\hbg$,
$Cl_{\psi,\dpsi}$ the Clifford algebra of auxiliary fermionic fields
$(\dpsi_{N+a},\psi_{N+a})$,
$Cl_{\tb,\tc}$ the Clifford algebra of fermionic ghosts
$(\tb_{i,j},\tc_{i,j})$,
$H_{\tB,\tG}$ the Heisenberg algebra of bosonic ghosts
$(\tB_{i,N+a},\tG_{i,N+a})$. The algebra of BRST cohomology
$H_{Q_{BRST}}({\cal A}_{tot})$ may be identified
with the quantum $W$ algebra.
According to Feigin-Frenkel \cite{FeFr}, we can simplify the full BRST complex
by decomposing the BRST currents into a constraint part associated with
an $sl(2)$ embedding and the canonical coboundary operator of the
nilpotent subalgebra.
This double complex can be analyzed by using
the spectral sequence technique.
Recently, a quite interesting observation was made by de Boer and Tjin.
They found that the components of non-trivial cohomology which
have zero gradation
in total degree of the double complex form
a closed algebra and the generators are nothing but a free field
realization of the quantum $W$-algebra\cite{BoTj}.
In presence of the second class constraints, however,
this standard decomposition
does not work in general \cite{SeTr}, which makes the analysis
more complicated.

In the present paper, we propose the following decomposition
$J_{BRST}(z)=J_{BRST}^{0}(z)+J_{BRST}^{1}(z)$, where
\eq
J_{BRST}^{1}(z)=-\sum_{i}\tc_{i,i+1}
-\sum_{a}\tc_{N+a,N+a+1}-\sum_{a}\tG_{N,N+a}\dpsi_{N+a},
\en
and $J_{BRST}^{0}(z)=J_{BRST}(z)-J_{BRST}^{1}(z)$.
Define two BRST charges $Q_{0}$ and $Q_{1}$ by contour integration of
$J_{BRST}^{0}(z)$ and $J_{BRST}^{1}(z)$, respectively.
Note that $J_{BRST}^0$ is different from the canonical
coboundary operator for the nilpotent subalgebra
by the term $\sum_{a<b} \tc_{N+a,N+b}\psi_{N+a}\dpsi_{N+b}$.
We can show that these BRST operators satisfy
\eq
Q_{0}^{2}=Q_{1}^{2}=Q_{0}Q_{1}+Q_{1}Q_{0}=0.
\en
It is convenient to introduce new currents
$\tJ_{i,j}$, $\tJ_{N+a,N+b}$, $\tj_{i,N+a}$ and $\tj_{N+a,i}$
(see Appendix A)
modified by the ghosts, which satisfy
\eqn
Q_{0}(\tb_{i,j})&=& \tJ_{i,j}, \quad i< j \CR
Q_{0}(-\tB_{i,N+a})&=& \tj_{i,N+a}, \CR
Q_{0}(\tb_{N+a,N+b})&=& \tJ_{N+a,N+b}, \quad a< b
\enn
The pairs $(\tJ_{i,j}, \tb_{i,j})$ ($i<j$),
$(\tJ_{N+a,N+b}, \tb_{N+a,N+b})$ ($a<b$) and
$(\tj_{i,N+a},-\tB_{i,N+a})$ form BRST doublets and decouple from the
non-trivial cohomology.
Therefore we consider only the reduced complex ${\cal A}_{red}$, which
spanned by other modified currents and ghosts.
In the following we shall take $N=3$;
the $sl(3|2)^{(1)}$ case for simplicity.

Let us consider the $Q_{0}$ action on the reduced complex.
The action on the modified currents is given by
\eqn
Q_{0}(\tJ_{2,1})&=& (\tH_{1}\tc_{1,2})+(K+1)\pa\tc_{1,2}, \CR
Q_{0}(\tJ_{3,2})&=& (\tH_{2}\tc_{2,3})+(K+1)\pa\tc_{2,3}
                    -(\tc_{4,5}(\psi_{4}\dpsi_{5})), \CR
Q_{0}(\tJ_{3,1})&=& ((\tH_{1}+\tH_{2})\tc_{1,3}) +(\tJ_{2,1}\tc_{2,3})
                     -(\tJ_{3,2}\tc_{1,2}) +(K+2)\pa\tc_{2,3}, \CR
Q_{0}(\tJ_{5,4})&=& -(\tH_{4}\tc_{4,5})-K \pa\tc_{4,5}, \CR
Q_{0}(\tj_{4,3})&=& (\tH_{3}\tG_{3,4})+(K+1)\pa\tG_{3,4}
                   +((\psi_{5}\dpsi_{5})\tG_{3,4})
                   -((\psi_{4}\dpsi_{5})\tG_{3,5}), \CR
Q_{0}(\tj_{4,2})&=& ((\tH_{2}+\tH_{4})\tG_{2,4})+(K+2)\pa\tG_{2,4}
                   +(\tJ_{3,2}\tG_{3,4})+(\tj_{4,3}\tc_{2,3})
                   -((\psi_{4}\dpsi_{4})\tG_{2,4})
                   -((\psi_{4}\dpsi_{5})\tG_{2,5}), \CR
Q_{0}(\tj_{4,1})&=& ((\tH_{1}+\tH_{2}+\tH_{3})\tG_{1,4}) +(K+3)\pa\tG_{1,4}
                   +(\tJ_{2,1}\tG_{2,4})+(\tJ_{3,1}\tG_{3,4}) \CR
& &                   +(\tj_{4,2}\tc_{1,2})+(\tj_{4,3}\tc_{1,3})
                   -((\psi_{4}\dpsi_{4})\tG_{1,4})
                   -((\psi_{4}\dpsi_{5})\tG_{1,5}), \CR
Q_{0}(\tj_{5,1})&=& ((\tH_{1}+\tH_{2}+\tH_{3}+\tH_{4})\tG_{1,5})
                    +(K+2)\pa\tG_{1,5}
                    +(\tJ_{2,1}\tG_{2,5}) +(\tJ_{3,1}\tG_{3,5}) \CR
& &                 +(\tj_{4,1}\tc_{4,5}) +(\tj_{5,2}\tc_{1,3})
                    -(\tJ_{5,4}\tG_{1,4})
                   -((\psi_{5}\dpsi_{4})\tG_{1,4})
                    -((\psi_{5}\dpsi_{5})\tG_{1,5}), \CR
Q_{0}(\tj_{5,2})&=& ((\tH_{2}+\tH_{3}+\tH_{4})\tG_{2,5})
                    +(K+1)\pa\tG_{2,5}
                    +(\tJ_{3,2}\tG_{3,5})-(\tJ_{5,4}\tG_{2,4}) \CR
& &                     +(\tj_{4,2}\tc_{4,5})+(\tj_{5,3}\tc_{2,3})
                    -((\psi_{5}\dpsi_{4})\tG_{2,4})
                    -((\psi_{5}\dpsi_{5})\tG_{2,5}), \CR
Q_{0}(\tj_{5,3})&=&((\tH_{2}+\tH_{3})\tG_{3,5})+K\pa\tG_{3,5}
                    +(\tj_{4,3}\tc_{4,5}) -(\tJ_{5,4}\tG_{3,4})
                   -((\psi_{5}\dpsi_{4})\tG_{3,4})
                   -((\psi_{5}\dpsi_{5})\tG_{3,5}).
\enn
Here we define the normal ordered product $(AB)(z)$ for two operators
$A(z)$ and $B(z)$ by
$(AB)(z)=\int_{z} {d w\over 2 \pi i} {A(w)B(z)\over w-z}$.
We take the $Q_{1}$-cohomology first. It is easy to see that
$H_{Q_{1}}({\cal A}_{red})$ is generated by
\eqn
\cU^{(0)}&=& 2 \tH_{1}+3\tH_{2}+6\tH_{3}+3\tH_{4}, \CR
\cG^{-\ (0)}&=& \dpsi_{5}, \CR
\cG^{+\ (0)}&=& \tj_{4,1}+\tj_{5,2}, \CR
\cT^{(0)}_{2}&=& \tJ_{2,1}+\tJ_{3,2}-\tJ_{5,4}+\tj_{5,3}\dpsi_{5}+
                  (\tj_{4,3}-\psi_{5})\dpsi_{4}, \CR
\cW^{(0)}_{2}&=& \tJ_{5,4}, \CR
\cW^{-\ (0)}&=& \dpsi_{4}, \CR
\cW^{+\ (0)}&=& \tj_{5,1}, \CR
\cW^{(0)}_{3}&=& \tJ_{3,1}+(\tj_{4,1}+\tj_{5,3})\dpsi_{4}+\tj_{4,3}\dpsi_{5}.
\enn
{}From the standard procedure of the spectral sequence \cite{BoTu},
we need to solve the descent equation
\eq
Q_{0}({\cal O}^{(n)})=Q_{1}({\cal O}^{(n+1)}).
\en
The generators of the total BRST cohomology are given by
\eq
{\cal O}={\cal O}^{(0)}-{\cal O}^{(1)}+{\cal O}^{(2)}+\cdots ,
\en
For a moment we shall restrict our concern to the $N=2$ superconformal sector.
The spin one operator $\cU^{(0)}$ satisfies $Q_{0}(\cU^{(0)})=0$.
We get $\cU=\cU^{(0)}$.
For a spin two operator $\cT^{(0)}$, we find $\cT=\cT^{(0)}-\cT^{(1)}$ where
\eqn
\cT^{(1)}&=& -{1\over2} (\tJ_{1,1}^{2}+\tJ_{2,2}^{2}+\tJ_{3,3}^{2}-
                       \tJ_{4,4}^{2}-\tJ_{5,5}^{2})
      -(\tH_{3}(\psi_{4}\dpsi_{4}))-((\tH_{3}+\tH_{4})(\psi_{5}\dpsi_{5})) \CR
& & -2 ((\psi_{4}\dpsi_{4})(\psi_{5}\dpsi_{5}))+((\tJ_{5,4}\psi_{4})\dpsi_{5}))
\CR
& &    -(\psi_{4}\pa\dpsi_{4})-(2+K) (\pa\psi_{4} \dpsi_{4})
 -(\psi_{5}\pa\dpsi_{5})-(1+K) (\pa\psi_{5}\dpsi_{5}) \CR
& &    -(1+K) \pa\tH_{1}-(2+K) \pa\tH_{3}-3 \tH_{3} +{K-1\over2}\pa\tH_{4} .
\enn
Next we consider the fermionic part.
For $\cG^{-\ (0)}$, we get $\cG^{-}=\cG^{-\ (0)}$.
For the operator $\cG^{+}$,
we find
$\cG^{+}=\cG^{+ \ (0)}-\cG^{+ \ (1)}+\cG^{+ \ (2)}-\cG^{+ \ (3)}$.
In the present paper we do not want to write down explicit formulae
for each term in the above expression due to complexity.
Compared to the $sl(2|1)^{(1)}$ case, all of the last terms
of solutions to the descent
equation cannot be written in terms of the modified
Cartan currents $\tH$'s and the auxiliary fermions.
We note that the argument of ref. \cite{BoTj} is not valid in this case,
because we have made a non-standard decomposition of the BRST current.
The gradation of the free fields in the double complex
depends on the way of decomposing the BRST operator.
Nevertheless we find that we can obtain a free field realization
$ O(z)$ of $N=2$ superconformal algebra by collecting
the terms which contain $\tH$'s and $\psi$, $\dpsi$'s from
the corresponding generator ${\cal O}(z)$ of the total
BRST cohomology.

We now derive the free field realization from the quantum
hamiltonian reduction described above.
Introduce four free bosons $\phi=(\phi_{1},\phi_{2},\phi_{3},\phi_{4})$ and
bosonize the modified Cartan currents $\tH_{i}=i a \A_{i}\cdot\pa\phi$
with $a=\sqrt{K+1}$.
In the following we use more explicit expressions:
\eqn
\tH_{1}&=&{i a \over \sqrt{2}}\pa\phi_{1}-i a \sqrt{3\over2}\pa\phi_{2}, \CR
\tH_{2}&=&{i a \over \sqrt{2}}\pa\phi_{1}+i a \sqrt{3\over2}\pa\phi_{2}, \CR
\tH_{3}&=&-{i a \over \sqrt{2}} \pa\phi_{1}-{i a\over \sqrt{6}}\pa\phi_{2}
          +{a\over \sqrt{6}}\pa\phi_{3} +{a\over\sqrt{2}}\pa\phi_{4}, \CR
\tH_{4}&=& -a \sqrt{2} \pa\phi_{4}.
\enn
The generators of $N=2$ superconformal algebras
are found to be ($v_{k}\equiv i \pa \phi_{k}$)
\begin{eqnarray}
U&=& (\psi_{4}\psi^{\dagger}_{4})+(\psi_{5}\psi^{\dagger}_{5})
     -i a\sqrt{6}v_{3}, \CR
G^{+}&=& G^{+}_{1}+G^{+}_{2}+ \pa(\DE G^{+}), \CR
G^{-}&=&\psi^{\dagger}_{5}, \CR
T^{N=2}&=& {1\over2} \sum_{i=1}^{4}v_{i}^{2}
    +\sqrt{2} (a-{1\over a}) \pa v_{1}
    -i \sqrt{3\over2}{1\over a} \pa v_{3}
    -i{1\over \sqrt{2}} (a+{1\over a})\pa v_{4} \CR
       & & -{3\over 2} (\pa \psi^{\dagger}_{4})\psi_{4}
           -{5\over 2} \psi^{\dagger}_{4} \pa \psi_{4}
           -{1\over 2} (\pa \psi^{\dagger}_{5})\psi_{5}
           -{3\over 2} \psi^{\dagger}_{5} \pa \psi_{5},
\end{eqnarray}
where
\eqn
G^{+}_{1}&=&\psi_{5}(T_{M}+T_{L}
  +{1\over2} T_{\psi^{\dagger}_{5},\psi_{5}}
  +T_{\psi^{\dagger}_{4},\psi_{4}} ),
\CR
G^{+}_{2}&=& \psi_{4} \left\{ (1+a^{2})(\pa^{2}\psi_{4}\dpsi_{4})
                            -{a \over \sqrt{3} b_{M}} W_{M}
                            -{2a \over\sqrt{6}}v_{+} T_{M}
                       +(a+{1\over a}) \pa T_{M}
          -\sqrt{6}a v_{+} (\pa\psi_{4}\dpsi_{4})\right.  \CR
      & & \left. +{a\over 3\sqrt{6}} \Bigl(
          4 i v_{+}^{3} +{6\sqrt{6}}(a+{1\over a})v_{+}\pa v_{+}
          -i {3 (2 a^{4}-a^{2}+2)\over a^{2}}
            \pa^{2}v_{+} \Bigr) \right\}
           , \CR
\DE G^{+}&=& \psi_{5}( i \sqrt{6} a v_{3}-(\psi_{4}\dpsi_{4}))
              +(1-3a^{2})\pa \psi_{5}
             + (1-2a^{2}) \psi_{4}(\pa\psi_{4}\dpsi_{4}) \CR
         & &  -a^{2}T_{M}\psi_{4}
            +i\sqrt{3\over2}\pa( v_{+}\psi_{4})
            +i \sqrt{6} a^{3} (\pa v_{+}) \psi_{4}
  +{(6a^{2}-1)(a^{2}-1)\over 6}\pa^{2}\psi_{4} .
\enn
Here we define the energy momentum tensors for $W_{3}$-minimal model (M),
$W_{3}$-gravity (L) and two ghost systems with spins $(3,-2)$ and $(2,-1)$;
\eqn
T_{M}&=& {1\over2} v_{1}^{2}+{1\over2} v_{2}^{2}
            +\sqrt{2} (a-{1\over a}) \pa v_{1}, \CR
T_{L}&=& {1\over2} v_{3}^{2}+{1\over2} v_{4}^{2}
         -i \sqrt{3\over2} (a+{1\over a})\pa v_{3}
         -i {1\over \sqrt{2}}(a+{1\over a})\pa v_{4}, \CR
T_{\psi^{\dagger}_{4},\psi_{4}} &=& -2 (\pa \psi^{\dagger}_{4})\psi_{4}
           -3 \psi^{\dagger}_{4} \pa \psi_{4}, \CR
T_{\psi^{\dagger}_{5},\psi_{5}} &=& - (\pa \psi^{\dagger}_{5})\psi_{5}
           -2 \psi^{\dagger}_{5} \pa \psi_{5}.
\enn
We also define a spin 3  current $W_{M}$
for the matter sector \cite{ZaFa}:
${6 i \over b_{M}} W_{M} =\sqrt{2} \tW_{M}$,
where
\eq
\tW_{M}= -i v_{2}^{3}-3 i v_{1}^{2}v_{2}
         -{3i \over \sqrt{2}}\A_{0}v_{2}\pa v_{1}
         -{9i \over \sqrt{2}}\A_{0}v_{1}\pa v_{2}
         -3i \A_{0}^{2} \pa^{2}v_{2},
\en
and $\A_{0}=a-{1\over a}$, $b_{M}={16 \over 22+ 5 c_{M}}$,
$c_{M}=2-24 \A_{0}^{2}$.
Then we have
\eqn
T^{N=2}+{1\over2} \pa U =T_{M}+T_{L}+T_{\psi^{\dagger}_{5},\psi_{5}}
+T_{\psi^{\dagger}_{4},\psi_{4}} .
\enn
The chiral supercurrent $G^{+}$ agrees with that used in
ref. \cite{BeBoRoTj} up to total derivative, after changing the variables
 $\phi_{1}\leftrightarrow\phi_{3}$,
 $\phi_{2}\leftrightarrow\phi_{4}$,
and $a\rightarrow i a$ .

The present realization looks different from \cite{BLNW}.
However if we transform the free fields by a similarity transformation
\cite{BeBoPaRo},
we can get the expression similar to ref. \cite{BLNW}.
Define a homotopy transformation $(A)^{R}(z)$ of an operator $A(z)$ by
\eq
(A)^{R}(z)\equiv e^{R}A(z) e^{-R}
=\sum_{i=0}^{\infty} {1\over i!}(A)^{R}_{i}(z),
\en
where $(A)^{R}_{i}=(\cR(A)^{R}_{i-1})_{-1}$, $(A)^{R}_{0}=A$ and
the generator is
\eq
R=\int {dz \over 2 \pi i}\cR (z) .
\en
$(AB)_{-1}(z)$ means the simple pole term in the OPE
between two operators $A(z)$ and $B(w)$.
We wish to choose $\cR$ such that the mixing term
$T_{M}v_{+}\psi_{4}$ does not appear in $(G^{+})^{R}$.
Then we find
\eq
\cR(z)=-{2i a\over \sqrt{6}}v_{+} (\psi_{4}\dpsi_{5})
\en
is such an operator.
This transformation was considered in ref. \cite{BeBoPaRo} in
the classical case.
Under the transformation, the free fields transform as
\eqn
(\psi_{5})^{R}&=& \psi_{5}-\sqrt{2\over3}a v_{+}\psi_{4}, \CR
(\dpsi_{4})^{R}&=& \dpsi_{4}+i \sqrt{2\over3}a v_{+} \dpsi_{5}, \CR
(v_{+})^{R} &=& v_{+}-i a^{2}\pa (\psi_{4}\dpsi_{5}).
\enn
Other fields are invariant.
Note that these fields are no longer free fields after this transformation.
In fact, there appear mixing terms in the $U(1)$ current and
the energy momentum tensor;
\eqn
U^{R} &=& U-a^{2}\pa(\psi_{4}\dpsi_{5}), \CR
(T^{N=2})^{R}&=& T^{N=2}-(1+{a^{2}\over2}) \pa^{2}(\psi_{4}\dpsi_{5}).
\enn
$G^{-}$ is invariant. $G^{+}$ transforms as
\eq
(G^{+})^{R}=G^{+}+(G^{+})^{R}_{1}+{1\over2} (G^{+})^{R}_{2}.
\en
Again $(G^{+})^{R}$ takes the form
\eq
(G^{+})^{R}=\cQ^{R}+\pa(\DE (G^{+})^{R}),
\en
where
\eqn
\cQ^{R}&=& \psi_{5}(T_{M}+T_{L}
  +{1\over2} T_{\psi^{\dagger}_{5},\psi_{5}}
  +T_{\psi^{\dagger}_{4},\psi_{4}} ) \CR
 & & +
    \psi_{4}\left(-{a\over \sqrt{3} b_{M}} W_{M}
                        +{1+a^{2}\over 2}\pa T_{M}
        +{i a\over \sqrt{3}b_{L}}W_{L} -{1+a^{2}\over 2}\pa T_{L}
                \right) \CR
& & +{a^{3} \over 3} (T_{M}-T_{L}) (\psi_{4}(\pa\psi_{4}\dpsi_{5})) \CR
& &  +a^{2} (\psi_{4}(\pa^{2}\psi_{5}\dpsi_{5}))
    -(\psi_{4}(\psi_{5}\pa^{2}\dpsi_{5}))
    -2 (\psi_{4}(\pa\psi_{5}\pa\dpsi_{5}))\CR
& &+  (-2 +{a^{2}\over2}+a^{4}) (\psi_{4}(\pa\psi_{4}\pa^{2}\dpsi_{5}))
   +(-{2\over3}-{a^{2}\over2}-{2a^{2}\over3})
   (\psi_{4}(\pa^{3}\psi_{4}\dpsi_{5}))  \CR
& &   +(-2-a^{2}-a^{4}) (\psi_{4}(\pa^{2}\psi_{4}\pa\dpsi_{5})) ,
\enn
\eqn
\DE (G^{+})^{R}
&=&  \psi_{5}( i \sqrt{6} a v_{3}-(\psi_{4}\dpsi_{4}))
              +(1-3a^{2})\pa \psi_{5} \CR
 & & -a^{2}T_{M}\psi_{4}
+\left(   {a^{2}\over2}v_{3}^{2}-{3a^{2}\over2}v_{4}^{2}
    +i\sqrt{3\over2} (a+a^{3})\pa v_{3}+{3i\over \sqrt{2}}\pa v_{4}\right)
     \psi_{4} \CR
& & -i \sqrt{2\over3} a^{3} v_{3}(\psi_{4}(\pa\psi_{4}\dpsi_{5}))
+2 (\psi_{4}(\psi_{5}\pa\dpsi_{5})) -a^{2} (\psi_{4}(\pa\psi_{5}\dpsi_{5}))
\CR
& & +({3 a^{2}\over2}+a^{4}) (\psi_{4}(\pa^{2}\psi_{4}\dpsi_{5})) .
\enn
We have introduced another spin $3$ current;
$
{6 i \over b_{L}} W_{L}= i \sqrt{2} \tW_{L}
$, where
\eq
\tW_{L}=-v_{3}^{3}+3 v_{3}v_{4}^{2}+3\sqrt{2} \B_{0} v_{3}\pa v_{4}
        +3\sqrt{3\over2}\B_{0} (v_{4}\pa v_{4} -v_3\pa v_{3})
        +{3\sqrt{3}\over2}\B_{0}^{2}\pa^{2}v_{4}
        -{3\over2}\B_{0}^{2}\pa^{2}v_{3}
\en
where $\B_{0}=-i(a+{1\over a})$, $b_{L}={16 \over 22+ 5 c_{L}}$,
$c_{L}=2-24 \B_{0}^{2}$.

Although the ghost coupling term is different from \cite{BLNW}, the ghost
fields coupled to the $W$-currents give the same expression.
But this comes from the fact that they use the same expression
of the energy momentum tensor as the untransformed one.

Next we argue the higher spin current part of
the topological $W$ algebra, although we do not
go into the detailed structure in the present paper.
{}From $\cW_{2}^{(0)}$ we get $\cW_{2}=\cW_{2}^{(0)}-\cW_{2}^{(1)}$.
We can read off the free field realization of another spin two current
$W_{2}(z)$ from $\cW_{2}$.
The result is
\eq
W_{2}(z)=-(\psi_{5}\dpsi_{4})
  +{1\over 4} (\tH_{4})^{2}+{1+K \over 2}\pa \tH_{4}.
\en
The $W_{2}$ is a quasi-primary field with respect to $T^{N=2}$.
But the current
\eq
\tW_{2}=W_{2}+{1+2 a^{2} \over 4(5-18 a^{2})} (UU)
        -{(1+2 a^{2})(1-3 a^{2}) \over 5-18 a^{2}} T^{N=2}
\en
becomes $N=2$ primary and obeys the OPE relation of $N=2$ super $W_{3}$
algebra:
\eq
\tW_{2}(z)\tW_{2}(w)={{1\over 2}\B^{2}c^{N=2}\over (z-w)^{4}}
                     +{A_{2}(w) \over (z-w)^{2}}
                     +{\pa A_{2}(w)\over z-w}+\cdots,
\en
where
\eq
A_{2}(z)={6 (1-5 a^{2})(1-2 a^{2}) \over 5-18 a^{2}} \tW_{2}
         +{2 \B^{2}c^{N=2} \over 5-18 a^{2}}
          \left( T^{N=2}-{1\over 4(1-3 a^{2})} (UU)\right)
\en
and $\B^{2}={(1+2a^{2}) (2-3 a^{2}) (1-4 a^{2}) \over 3 (5-18 a^{2})}$.
Other currents in the $N=2$ multiplet are defined
by $\tW^{+}=G^{+}_{-1/2}\tW_{2}$,
$\tW^{-}=G^{-}_{-1/2}\tW_{2}$ and $\tW_{3}=G^{+}_{-1/2}G^{-}_{-1/2}\tW^{2}$.
It is a straightforward task to check that they form the $N=2$ super-$W_{3}$
algebra.

In the present paper we have established a new free field realization of
$N=2$ super-$W_{3}$ algebra, which cannot be written in superspace
formalism.
To study the degenerate representation of
the $N=2$ super-$W_{3}$ algebra, we need the
screening operators associated with the simple roots $\A_{i}$ of $\bg$
\eqn
S^{+}_{\A_{j}}&=& e^{i a \A_{j}\cdot \phi}, \quad
S^{-}_{\A_{j}}=e^{-i {1\over a} \A_{j}\cdot \phi}, \quad (j=1,2), \CR
S^{+}_{\A_{4}}&=& e^{-i a \A_{4}\cdot \phi}, \quad
S^{-}_{\A_{4}}= e^{-i {1\over a} \A_{4}\cdot \phi},\CR
S_{\A_{3}}(z)&=& \dpsi_{4} e^{-{i\over a}\A_{3}\cdot \phi}.
\enn
The singular vectors in the Fock modules of bosons
$\A_{i}\cdot\phi$ for $i=1,2$ and $4$ can be characterized by using the
associated screening operators.
We take the irreducible representation for these bosons.
There is a single boson $\lm_{3}\cdot\phi=\phi_{3}$
which is orthogonal to these
bosons, where $\lm_{i}$ denotes the fundamental weight of $\bg$:
$\A_{i}\cdot\lm_{j}=\D_{i,j}$. Contrary to the other bosons,
the Hilbert space of $\phi_{3}$ remains to be the Fock module.
In the case of $sl(2|1)^{(1)}$, the corresponding free field
is identified as the Liouville
field, which plays an essential role in the analysis of the BRST
cohomology of non-critical string theory.
One might expect that $\phi_{3}$ plays
a similar role in the $W$-string theory.
But this cannot be the end of story.
This $N=2$ model would effectively reduce to a model of
non-critical string theory, if only the Liouville dressing was
considered. Therefore, if we want to introduce $W$-dressing,
which is indispensable to $W$ gravity coupling, we need
one more Fock module of a free boson. This can be achieved,
for example, by ignoring the screening operator $S^{\pm}_{\A_{4}}$
in the degenerate representation.
This suggests that the representation theory of a non-compact
Lie superalgebra $sl(3|2)$ with the bosonic subalgebra
$sl(3)\oplus sl(2, {\bf R})\oplus u(1)$
is preferable for the investigation of the non-trivial
observable in the non-critical $W_{3}$-string theory.
It would be interesting to examine the recent analysis
of the physical spectrum of the non-critical
$W_3$-string\cite{BMP}\cite{BeBoRoTj} from such a viewpoint of
Lie superalgebra.

\vspace{1cm}

The work of K.I. is partially supported by University of Tsukuba Reserach
Projects.

\newpage

{\bf Appendix A: Definition for currents modified by ghost fields}
\begin{eqnarray*}
\tJ_{i,j}(z) &=&\left\{
\begin{array}{ll}
                J_{i,j}-\sum_{k=j+1}^{N}\tc_{j,k}\tb_{i,k}
                +\sum_{k=1}^{i-1}\tc_{k,i}\tb_{k,j}-
                \sum_{a=1}^{N-1}\tG_{j,N+a}\tB_{N+a},
& \mbox{for $i<j$,} \\
& \\
              J_{i,i}+\sum_{k=1}^{i-1}\tc_{k,i}\tb_{k,i}
                       -\sum_{k=i+1}^{N}\tc_{i,k}\tb_{i,k}
                    -\sum_{a=1}^{N-1}\tG_{i,N+a}\tB_{i,N+a} & \\
               \quad        -\D_{i,N}\sum_{a=1}^{N-1}\psi_{N+a}\dpsi_{N+a},
                 & \mbox{for $i=j$,} \\
& \\
J_{i,j}+\sum_{k=1}^{i-1}\tc_{k,i}\tb_{k,j}
                   -\sum_{k=j+1}^{N}\tc_{j,k}\tb_{i,k}
                   -\sum_{a=1}^{N-1}\tG_{j,N+a}\tB_{i,N+a}, &
                 \mbox{for $i>j$}.
               \end{array}
\right.  \CR
\tJ_{N+a,N+b}(z)&=& \left\{
\begin{array}{ll}
J_{N+a,N+b}-\sum_{i=1}^{N}\tG_{i,N+a}\tB_{i,N+b}
+\sum_{c=b+1}^{N-1}\tc_{N+b,N+c}\tb_{N+a,N+c} & \\
                     -\sum_{c=1}^{a-1}\tc_{N+c,N+a}\tb_{N+d,N+b}
                    -\psi_{N+a}\dpsi_{N+b}, & \mbox{ for $a<b$} \\
& \\
J_{N+a,N+a} -\sum_{i=1}^{N}\tG_{i,N+a}\tB_{i,N+a}
                       -\psi_{N+a}\dpsi_{N+a} & \\
-\sum_{c=1}^{a-1}\tc_{N+c,N+a}\tb_{N+c,N+a}
                +\sum_{c=a+1}^{N-1}\tc_{N+a,N+c}\tb_{N+a,N+c}, &
\mbox{ for $a=b$} \\
& \\
J_{N+a,N+b}-\sum_{i=1}^{N}\tG_{i,N+a}\tB_{i,N+b}
                       -\sum_{c=1}^{b-1}\tc_{N+c,N+a}\tb_{N+c,N+b} & \\
+\sum_{c=a+1}^{N-1}\tc_{N+b,N+c}\tb_{N+a,N+c}
                       +\dpsi_{N+b}\psi_{N+a}, & \mbox{ for $a>b$}.
\end{array}
\right. \CR
\tj_{i,N+a}(z)&=&j_{i,N+a}-\sum_{k=1}^{i-1}\tc_{k,i}\tB_{k,N+a}
                    -\sum_{b=a+1}^{N-1}\tc_{N+a,N+b}\tB_{i,N+b}, \CR
\tj_{N+a,i}(z)&=& \tj_{N+a,i}-\sum_{k=1}^{i-1}\tG_{k,N+a}\tb_{k,i}
                    +\sum_{b=a+1}^{N-1}\tG_{i,N+b}\tb_{N+a,N+b}. \nonumber
\end{eqnarray*}

\end{document}